\documentclass[aps,pra,twocolumn,showpacs,superscriptaddress]{revtex4-1}
\usepackage{amssymb}
\usepackage{amsmath}
\usepackage{amsthm}
\usepackage{txfonts}
\usepackage{graphicx}
\usepackage[colorlinks, linkcolor=blue, citecolor=blue, urlcolor=blue]{hyperref}

\begin{document}

\title{Lower bound of local quantum uncertainty for high-dimensional bipartite quantum systems}

\emph{}
\author{Shuhao Wang}
\affiliation{State Key Laboratory of Low-Dimensional Quantum Physics and Department of Physics, Tsinghua University, Beijing 100084, China}
\author{Hui Li}
\affiliation{School of Science, Tianjin University of Technology, Tianjin 300191, China}
\author{Xian Lu}
\affiliation{Institute of Software, Chinese Academy of Sciences, Beijing 100080, China}
\author{Bin Chen}
\affiliation{State Key Laboratory of Low-Dimensional Quantum Physics and Department of Physics, Tsinghua University, Beijing 100084, China}
\author{Gui-Lu Long}
\email{gllong@tsinghua.edu.cn}
\affiliation{State Key Laboratory of Low-Dimensional Quantum Physics and Department of Physics, Tsinghua University, Beijing 100084, China}
\affiliation{Tsinghua National Laboratory for Information Science
and Technology, Beijing 100084, China}
\affiliation{Collaborative Innovation Center of Quantum Matter, Beijing 100084, China}
\date{\today }

\begin{abstract}
Quantum correlations are of fundamental importance in quantum phenomena and quantum information processing studies. The measure of quantum correlations is one central issue. The recently proposed measure of quantum correlations, the local quantum uncertainty (LQU), satisfies the full physical requirements of a measure of quantum correlations. In this work, by using operator relaxation, a closed form lower bound of the LQU for arbitrary-dimensional bipartite quantum states is derived. We have compared the lower bound and the optimized LQU for several typical quantum states.
\end{abstract}


\maketitle

\section{Introduction}
Over the past decades, entanglement was considered to be the only ingredient of quantum properties and the main resource of the speed-up in quantum computation \cite{epr,schrodinger,rmp-81-865}. However, it has been shown that some states without entanglement but with quantum correlations as measured by quantum discord can still reveal their power in quantum speed-up \cite{prl-88-017901,prl-81-5672,jpa-34-6899}. Nowadays, it is widely believed that the non-classical correlations, namely, quantum correlations, play vital roles in the quantum features in quantum information processing. The investigation of quantum correlations is of fundamental importance in the study of quantum phenomena in nature.
As a result, quantum correlations become the subject of intensive studies in the last two decades \cite{rmp-84-1655}. Among varies researches, it is of great significance to measure quantum correlations quantitatively. There are much attention put on the measurement of bipartite quantum correlations, including quantum discord \cite{prl-88-017901,jpa-34-6899,zurek2013}, geometric discord \cite{prl-105-190502,pra-86-012312}, quantum deficit \cite{pra-66-022104}, measurement-induced disturbance \cite{pra-77-022301}, etc.

For high-dimensional bipartite quantum states without high-symmetry, it is considerably hard to avoid the optimization in the calculation of the quantum discord \cite{chen2011,davide2015}. Actually, it is widely accepted that the calculation of quantum discord is NP-hard \cite{huang2015}. Therefore people have contributed a lot of attention to finding the lower bound of various types of quantum correlation definitions \cite{luo2010,hassan2012}.

Recently, a measure of quantum correlations for bipartite quantum systems named the local quantum uncertainty (LQU) is proposed \cite{prl-110-240402}. The LQU is defined as
\begin{equation}
\mathcal{U}_A^\Lambda=\min_{K^\Lambda} I(\rho_{AB}, K^\Lambda),
\end{equation}
where we have denoted the two particles as $A$ and $B$, the minimum is optimized over all the non-degenerate observables on A: $K = K_A^\Lambda \otimes \mathbb{I}_B$, and
\begin{equation}
I(\rho, K)= -\frac{1}{2}{\rm Tr}([\sqrt{\rho},K]^2)
\end{equation}
is the skew information \cite{wigner}, where $[{\cdot},{\cdot}]$ denotes the commutator. It has been shown that for bipartite quantum systems, the LQU is invariant under local unitary operations, non-increasing under local operations on $B$, vanishes if and only if the quantum state is a zero discord state with respect to measurements on $A$. For pure states, the LQU is an entanglement monotone. In a word, the LQU satisfies the full physical requirements of a measure of quantum correlations \cite{prl-110-240402}.

The advantage of the LQU over quantum discord lies on the possibility of to obtaining the closed form.
The closed form of the LQU for $2 \times d$ quantum systems  \cite{prl-110-240402} is pointed out to be
\begin{equation}
\label{twod}
\mathcal{U}_A=1-\lambda_{\rm max}(\mathcal{W}),
\end{equation}
where $\lambda_{\rm max}$ is the maximum eigenvalue of the $3 \times 3$ matrix $\mathcal{W}$ with elements $\mathcal{W}_{ij}={\rm Tr}\{\sqrt{\rho}(\sigma_{i}\otimes \mathbb{I})\sqrt{\rho}(\sigma_{j}\otimes \mathbb{I})\}$, and $\sigma_{i}$ $(i=1,2,3)$ represent the Pauli matrices, which are the generators of ${\rm SU}(2)$ (the special unitary group of degree 2).
The interesting coincidence arises that for $2 \times d$ quantum systems, the LQU reduces to the linear entropy (i.e., the concurrence) for pure states.
However, for high-dimensional quantum systems, it is still a hard bone to obtain the LQU without cumbersome optimization.
A ray of hope comes from \cite{arxiv}, in which the authors pointed out that the closed form of the LQU can be achieved for $d \times d$ quantum states with high symmetry with a relaxation of the operators in the optimization.

In this paper, we seek for the possibility to obtain a closed form lower bound of LQU for high-dimensional quantum systems. We achieve our goal by using the same operator relaxation approach as in \cite{arxiv}. Several representative quantum states are studied by comparing our lower bound and optimized LQU obtained by genetic algorithm.

\section{Preliminaries}

There exists one important requirement on the optimization operators in the definition of the LQU, namely, they should have non-degenerate fixed spectrum. Suppose the non-degenerate fixed spectrum is chosen as $\Lambda$, the operators with this spectrum in subspace $A$ thus can be parametrized by
\begin{equation}
\label{operatorsu}
K_A^\Lambda = V_A \Lambda V_A^\dagger,
\end{equation}
where $V_A$ varies over the special unitary group on $A$.

The SU($d$) group can be decomposed into the product of $d(d-1)/2$ basic transformations with $d^2 - 1$ parameters using Hurwiz's theory \cite{huili2013}. By indicating the LQU as the objective function and optimizing over these parameters, we can get its accurate value with computational methods such as genetic algorithm.

Calculating the LQU by using computational approaches is definitely complex for real-world applications. In the following, we revisit the derivation of the closed form of the LQU for $2 \times d$ quantum systems. A qubit observable with non-degenerate fixed spectrum $\sigma_z$ can be parametrized by \cite{prl-110-240402}
\begin{equation}
\label{extendtwo}
K_A = \alpha \vec{s} \cdot \vec{\sigma} + \beta \mathbb{I},
\end{equation}
where $|\vec{s}| = 1$ and $\alpha \ne 0$.
Therefore in this case, with different values of $\alpha$, the LQU is equivalent to
\begin{equation}
\label{twodLQU}
\mathcal{U}_A=\min_{\vec{s}} I(\rho_{AB}, \vec{s} \cdot \vec{\sigma} \otimes \mathbb{I}_d)
\end{equation}
up to a constant multiplier $|\alpha|^2$.

Eq. (\ref{twodLQU}) gives us the possibility to obtain the closed form
\begin{eqnarray}
\mathcal{U}_A&=&\min_{\vec{s}} I(\rho_{AB}, \vec{s} \cdot \vec{\sigma} \otimes \mathbb{I}_d)\nonumber\\
& = & \min\sum_{i,j} s_i s_j [{\rm Tr}\{\rho_{AB} \sigma_i \sigma_j - \sqrt{\rho}(\sigma_i \otimes \mathbb{I}_d) \sqrt{\rho}(\sigma_j \otimes \mathbb{I}_d)\}]\nonumber\\
& = & 1 - \min\sum_{i,j} s_i s_j {\rm Tr}\{\sqrt{\rho}(\sigma_i \otimes \mathbb{I}_d) \sqrt{\rho}(\sigma_j \otimes \mathbb{I}_d)\}.
\end{eqnarray}
This optimization gives Eq. (\ref{twod}).

\section{Lower bound of LQU}

For $d_1 \times d_2$ quantum states, the key to calculating the LQU is the optimization among operators with non-degenerate fixed spectrum on one party of the quantum system, say $A$. The difficulty lies in the parametrization of these operators. The construction of qubit operators gives us the sign to this problem.

Similar to qubit operators, higher-dimensional operators can be expressed as
\begin{equation}
K_A^\alpha=\vec{s}\cdot \vec{\lambda} + \beta \mathbb{I}_{d_1},
\end{equation}
where $\vec{s}=(s_1,s_2,...,s_{d_1^2-1})$ and $|\vec{s}|=\alpha \ne 0$, $\lambda=(\lambda_1,\, \lambda_2,...,\lambda_{d_1^2-1})^T$ is the vector formed by the generators of SU($d_1$) group. Note that this expression is slightly different from Eq. (\ref{extendtwo}).
Different from qubit operators, higher-dimensional operators can not easily satisfy the non-degenerate fixed spectrum requirement.

One importation observation is the following theorem.

{\bf Theorem 1.} The operators with the same non-degenerate fixed spectrum belong to the same set $K_A^\alpha$.

{\bf Proof.} Suppose we have a non-degenerate fixed spectrum $\Lambda$, which can be extended as
\begin{equation}
\Lambda = \vec{s}\cdot \vec{\lambda} + \beta \mathbb{I}_{d_1},
\end{equation}
where $|\vec{s}|=\alpha$.

Following Eq. (\ref{operatorsu}), we can also extend the operator after unitary transformation
\begin{equation}
K_A^\Lambda = \vec{s'}\cdot \vec{\lambda} + \beta' \mathbb{I}_{d_1}.
\end{equation}

By using the fact that
\begin{eqnarray}
&&{\rm Tr} K_A^\Lambda = {\rm Tr} \{V_A \Lambda V_A^\dagger\} = {\rm Tr}\Lambda\nonumber\\
&&{\rm Tr}(K_A^\Lambda)^2 = {\rm Tr} \Lambda^2,
\end{eqnarray}
we get $\beta = \beta'$, $|\vec{s}| = |\vec{s'}|$.

Therefore, $K_A^\Lambda$ also belongs to the set $K_A^\alpha$.\qed

Based on this observation, the lower bound is possible to obtain by using operator relaxation, namely, we do not require operators with non-degenerate fixed spectrum. After choosing the spectrum, the only work we need to do before optimization is determining $\alpha$ by extending the spectrum with $\vec{\lambda}$ and $\mathbb{I}$.

Reminding that there are $d^2-1$ generators of SU($d$) denoted as
\begin{equation}
\label{generators}
\lambda_{j} =
\begin{cases}
\sqrt{\frac{2}{j(j+1)}}\left(\sum_{k\,=\,1}^{j}|k\rangle\langle k|-j|j+1\rangle\langle j+1|\right),j=1,...,d-1\\
|k\rangle\langle m|+|m\rangle\langle k| (1\leq k<m\leq d), j=d,...,\frac{d(d+1)}{2}-1\\
\mathrm{i}( |k\rangle\langle m|-|m\rangle\langle k|) (1\leq k<m\leq d), j=\frac{d(d+1)}{2},...,d^2-1\\
\end{cases}.
\end{equation}
They satisfy
\begin{equation}\label{pro}
\lambda_i \lambda_j=\mathrm{i} \sum_k{f_{ijk} \lambda_k}+ \sum_k{g_{ijk} \lambda_k}+\frac{2}{d}\delta_{ij}\mathbb{I}_d,
\end{equation}
where
\begin{eqnarray}
f_{ijk}=\frac{1}{4\mathrm{i}}{\rm Tr}([\lambda_i,\lambda_j]\lambda_k),
g_{ijk}=\frac{1}{4}{\rm Tr}(\{\lambda_i,\lambda_j\}\lambda_k),
\end{eqnarray}
where $\{\cdot,\cdot\}$ represents the anti-commutator.

{\bf Theorem 2.} The closed form lower bound of the LQU for $d_1 \times d_2$ quantum states is
\begin{eqnarray}
\label{CF}
\mathcal{U}_A = \alpha^2 (\frac{2}{d_1} - \lambda_{\rm max}(\mathcal{W})),
\end{eqnarray}
where we have used $\lambda_{\rm max}$ to represent the maximum eigenvalue, $\mathcal{W}$ is a $(d_1^2-1)\times(d_1^2-1)$ matrix with elements
\begin{equation}
\label{W}
\mathcal{W}_{ij}={\rm Tr}\{\sqrt{\rho}(\lambda_{i}\otimes \mathbb{I}_{d_2})\sqrt{\rho}(\lambda_{j}\otimes \mathbb{I}_{d_2})\}-G_{ij} L,
\end{equation}
and
\begin{eqnarray}
&&G_{ij} = (g_{ij1}, \cdots , g_{ijk}, \cdots , g_{ijd_1^2-1}),\nonumber\\
&&L = ({\rm Tr}(\rho \lambda_{1}\otimes \mathbb{I}_{d_2}), \cdots , {\rm Tr}(\rho \lambda_{k}\otimes \mathbb{I}_{d_2}), \nonumber\\
&&\cdots , {\rm Tr}(\rho \lambda_{d_1^2-1}\otimes \mathbb{I}_{d_2}))^T.
\end{eqnarray}

{\bf Proof.} Following the definition of the LQU, we obtain \cite{luo2003}
\begin{eqnarray}
\mathcal{U}_A&=& \min{I(\rho, K)}\nonumber\\
&=& \min{\{{\rm Tr}(\rho (K)^2)-{\rm Tr}(\sqrt{\rho} K \sqrt{\rho} K)}\}\nonumber\\
&=& \min\{{\rm Tr}\{\rho(\vec{s}\cdot\lambda\otimes \mathbb{I}_{d_2})^2\}\nonumber\\
&&-{\rm Tr}\{\sqrt{\rho} (\vec{s}\cdot\lambda\otimes \mathbb{I}_{d_2}) \sqrt{\rho}(\vec{s}\cdot\lambda\otimes \mathbb{I}_{d_2})\}\}.
\end{eqnarray}
By using Eq. (\ref{pro}), it is easy to get
\begin{eqnarray}
&&{\rm Tr}\{\rho(\vec{s}\cdot\lambda\otimes \mathbb{I}_{d_2})^2\}=\nonumber\\
&&\sum_{i,j,k} s_i s_j[(\mathrm{i} f_{ijk}+g_{ijk}){\rm Tr}(\rho \lambda_{k}\otimes \mathbb{I}_{d_2})]+\frac{2\alpha^2}{d_1}.
\end{eqnarray}
We define
\begin{equation}
F_{ij}=(f_{ij1}, \cdots , f_{ijk}, \cdots , f_{ijd_1^2-1}),
\end{equation}
Then
\begin{eqnarray}
{\rm Tr}\{\rho(\vec{s}\cdot\lambda\otimes \mathbb{I}_{d_2})^2\}=
\sum_{i,j} s_i s_j[(\mathrm{i} F_{ij}+G_{ij})L]
 +\frac{2\alpha^2}{d_1}.
\end{eqnarray}
Thus we get
\begin{eqnarray}
\mathcal{U}_A&=& \frac{2\alpha^2}{d_1} + \min\sum_{i,j} s_i s_j[(\mathrm{i} F_{ij}+G_{ij})L\nonumber\\
&&-{\rm Tr}\{\sqrt{\rho}(\lambda_{i}\otimes \mathbb{I}_{d_2})\sqrt{\rho}(\lambda_{j}\otimes \mathbb{I}_{d_2})\}].
\end{eqnarray}
It can be seen that $F_{ij}$ is antisymmetric under the transpose of the subscripts, namely, $F_{ji}=-F_{ij}$, therefore $\sum_{i,j} s_i s_j F_{ij} = 0$.

Therefore, we finally get
\begin{equation}
\mathcal{U}_A = \frac{2\alpha^2}{d_1} + \min\sum_{i,j} s_i s_j[ G_{ij} L-{\rm Tr}\{\sqrt{\rho}(\lambda_{i}\otimes \mathbb{I}_{d_2})\sqrt{\rho}(\lambda_{j}\otimes \mathbb{I}_{d_2})\}].
\end{equation}
This optimization arrive at the closed form given in Eq. (\ref{CF}).\qed

Two special cases are worthwhile to be discussed. In the case where $d_1=2$, we have $g_{ijk}=0$, thus $G_{ij}$ is a zero vector. It is easy to recover the result in Eq. (\ref{twod}). When ${\rm Tr}(\rho \lambda_i \otimes \mathbb{I}_d)=0$, namely, $L=0$, the conclusion in \cite{arxiv} is recovered.

\section{Lower Bound vs. Optimized LQU}

In the section, we study the lower bound and the optimized LQU of two kinds of quantum systems, i.e., qutrit-qutrit and qudit-qubit states. The optimized LQU is obtained by minimizing the skew information within non-degenerate operators with a fixed spectrum using genetic algorithm.

\subsection{Qutrit-qutrit states}

In this case, we choose the non-degenerate fixed spectrum as
\begin{equation}
\Lambda = \left( {\begin{array}{*{20}{c}}
1&0&0\\
0&-1&0\\
0&0&0
\end{array}} \right).
\end{equation}
It can been easily verified from Eq. (\ref{generators}) that $\Lambda = \lambda_1$. Therefore, we have $\alpha = |\vec{s}| = 1$.

We first consider the Werner state as an example. The qutrit-qutrit Werner state is defined as \cite{Werner}
\begin{eqnarray}
\rho_{w} = p \left| \psi  \right\rangle \langle \psi | + \frac{(1-p)}{9}{\mathbb{I}_3},
\end{eqnarray}
where $\left| \psi  \right\rangle = \frac{1}{{\sqrt 3 }}\sum\limits_{i = 0}^2 {{{\left| i \right\rangle }_1}} {\left| i \right\rangle _2}$ and $ 0 \le p \le 1$. The Werner state is is highly symmetric \cite{eric} and ${\rm Tr}(\rho_w \lambda_i \otimes \mathbb{I}_3)=0$.

\begin{figure}[tbp]
\begin{center}
\includegraphics[width=8cm]{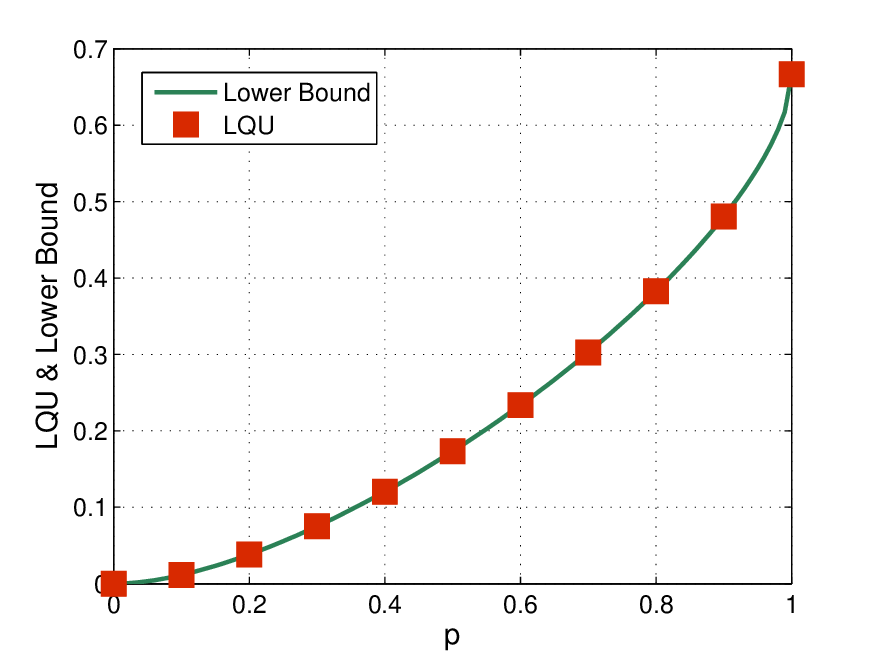}
\end{center}
\caption{The LQU of the qutrit-qutrit Werner state.} \label{Werner}
\end{figure}

From Fig. \ref{Werner} we see that when $p=0$, the quantum system only exists white noise, in this case the LQU is zero. While one increases $p$, the LQU increases. Therefore, the LQU is maximized in the case where $p=1$.

According to \cite{arxiv}, the Werner state is isotropic, thus the lower bound and the optimized LQU are identical.

\begin{figure}[tbp]
\begin{center}
\includegraphics[width=8cm]{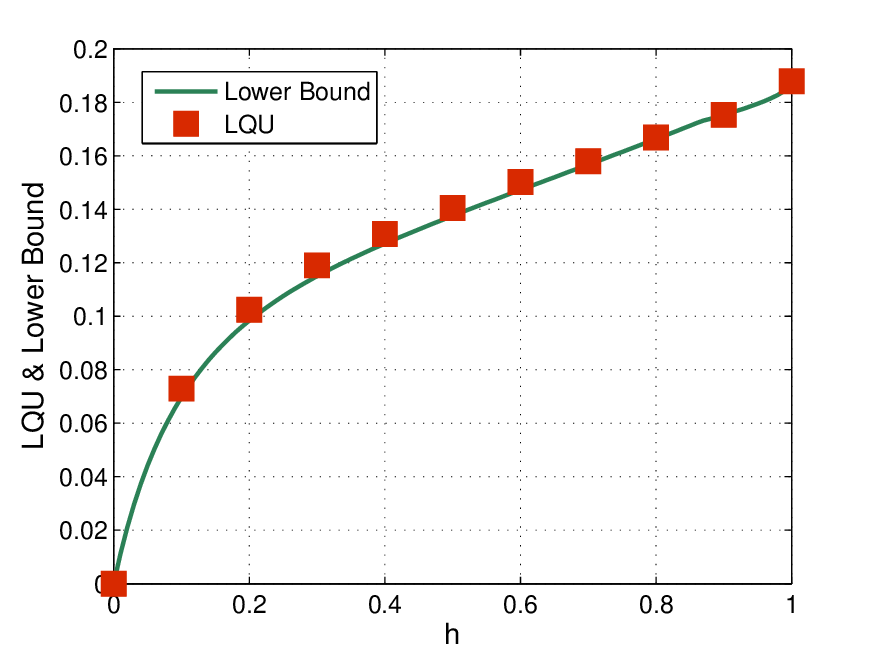}
\end{center}
\caption{The LQU of the qutrit-qutrit Horodecki state.} \label{Horo}
\end{figure}

Then we investigated the qutrit-qutrit Horodecki state \cite{horo42}
\begin{eqnarray}
{\rho_h} = \frac{1}{{8h + 1}}\left( {\begin{array}{*{20}{c}}
h&0&0&0&h&0&0&0&h\\
0&h&0&0&0&0&0&0&0\\
0&0&h&0&0&0&0&0&0\\
0&0&0&h&0&0&0&0&0\\
h&0&0&0&h&0&0&0&h\\
0&0&0&0&0&h&0&0&0\\
0&0&0&0&0&0&{\frac{{(1 + h)}}{2}}&0&{\frac{{\sqrt {1 - {h^2}} }}{2}}\\
0&0&0&0&0&0&0&h&0\\
h&0&0&0&h&0&{\frac{{\sqrt {1 - {h^2}} }}{2}}&0&{\frac{{(1 + h)}}{2}}
\end{array}} \right).
\end{eqnarray}
It is a partial positive transpose (PPT) entangled state. Different from the qutrit-qutrit Werner state, the qutrit-qutrit Horodecki state is not symmetric and does not satisfy ${\rm Tr}(\rho_h \lambda_i \otimes \mathbb{I}_3)=0$.
Although it is a PPT state, it can be seen from Fig. \ref{Horo} that the LQU is non-zero when $h > 0$. Meanwhile, the lower bound is tight respect to the optimized LQU.

As the last example for the qutrit-qutrit case, we study the LQU of the generalized qutrit-qutrit Bell state defined as
\begin{equation}
\mathcal{\rho}_{AB}(0)=(|00\rangle+|11\rangle+|22\rangle)(\langle 00|+\langle 11|+\langle 22|)
\end{equation}
during decoherence.

Reminding that the docoherence process can be written in terms of Kraus operators as
\begin{equation}
\mathcal{\rho}_{AB}(t)=\sum_{j=1}^3\sum_{i=1}^3 \mathcal{F}_j^B \mathcal{E}_i^A \mathcal{\rho}_{AB}(0){\mathcal{E}_i^{A}}^\dagger {\mathcal{F}_j^{B}}^\dagger,
\end{equation}
where the operators $\mathcal{E}_i^A$ and $\mathcal{F}_j^B$ are the Kraus operators describing the noise channels on particles
$A$ and $B$, they satisfy $ \sum_{i}\mathcal{E}_i^A{\mathcal{E}_i^{A}}^\dagger=\mathbb{I} $ and $ \sum_i\mathcal{F}_i^B{\mathcal{F}_i^{B}}^\dagger=\mathbb{I} $.

We impose two dephasing channels on particles $A$ and $B$, the Kraus operators of the dephasing channel, for instance, on particle $A$ are
\begin{eqnarray}
&&\mathcal{E}_1^A=\begin{pmatrix}
  1 & 0 & 0 \\
  0 & \sqrt{1-\gamma_A} & 0 \\
  0  & 0  &  \sqrt{1-\gamma_A}  \\
  \end{pmatrix} \otimes\mathbb{I}_3\nonumber\\
&&\mathcal{E}_2^A=\begin{pmatrix}
  0 & 0 & 0 \\
  0 & \sqrt{\gamma_A} & 0 \\
  0  & 0  & 0  \\
  \end{pmatrix} \otimes\mathbb{I}_3\nonumber\\
&&\mathcal{E}_3^A=\begin{pmatrix}
  0 & 0 & 0 \\
  0 & 0 & 0 \\
  0  & 0  & \sqrt{\gamma_A}  \\
  \end{pmatrix}\otimes\mathbb{I}_3
\end{eqnarray}
where $\gamma_A$ denotes the dephasing strength.

\begin{figure}[tbp]
\begin{center}
\includegraphics[width=8cm]{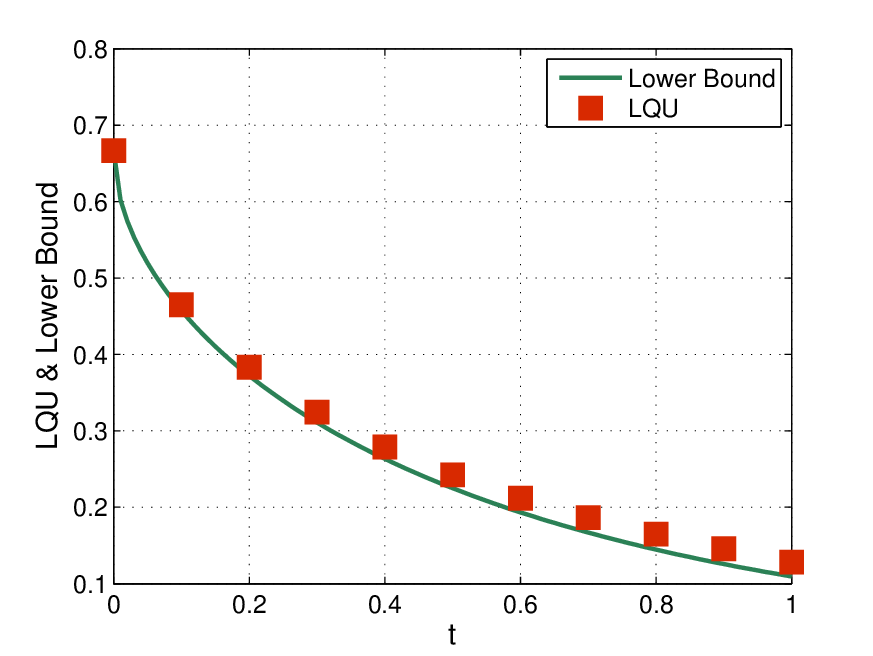}
\includegraphics[width=8cm]{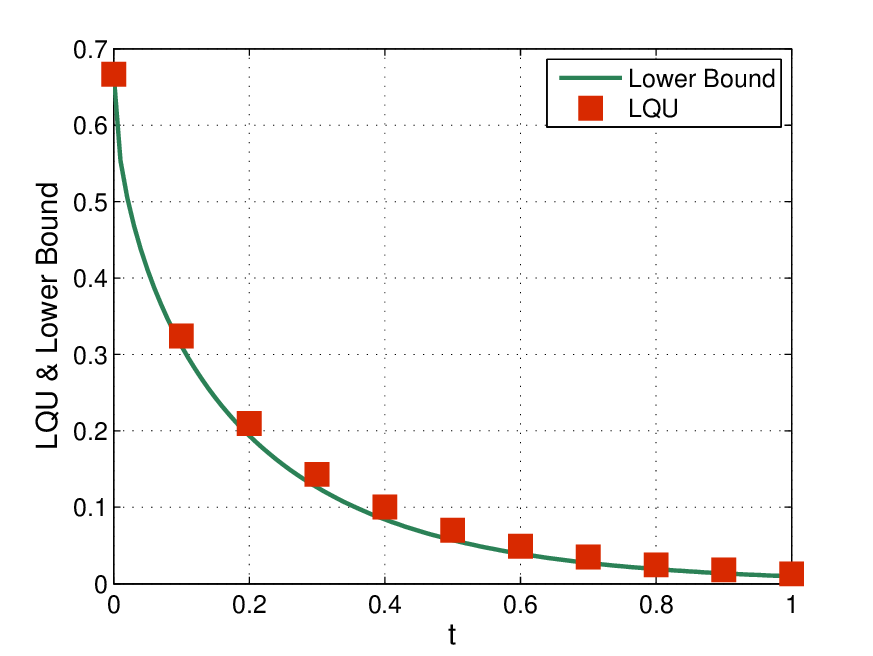}
\end{center}
\caption{The LQU of the generalized qutrit-qutrit Bell state going through two dephasing channels with (Above) $\gamma_A = \gamma_B = 0.5$; (Bottom) $\gamma_A = 2.0, \gamma_B = 1.0$.}
\label{decoherence}
\end{figure}

We have investigated two situations where $\gamma_A = \gamma_B = 0.5$ and $\gamma_A = 2.0, \gamma_B = 1.0$, respectively. The lower bound and the optimized LQU for these two cases are shown in Fig. \ref{decoherence}. The trend of the lower bound is the same as the optimized LQU, and the bound is close to the optimized value.

\subsection{A qudit-qubit state}

As a much more complicated example, we consider the case where the dimension of particle $A$ is four. The non-degenerate fixed spectrum is chosen as
\begin{equation}
\Lambda = \sqrt{\frac{2}{3}}\left( {\begin{array}{*{20}{c}}
3&0&0&0\\
0&1&0&0\\
0&0&-1&0\\
0&0&0&-3
\end{array}} \right).
\end{equation}
We can verify that $\Lambda = \sqrt{\frac{2}{3}}(\lambda_1+\sqrt{\frac{1}{3}}\lambda_2+\sqrt{\frac{1}{6}}\lambda_3)$. In this case, we also have $\alpha = |\vec{s}| = 1$.

\begin{figure}[tbp]
\begin{center}
\includegraphics[width=8cm]{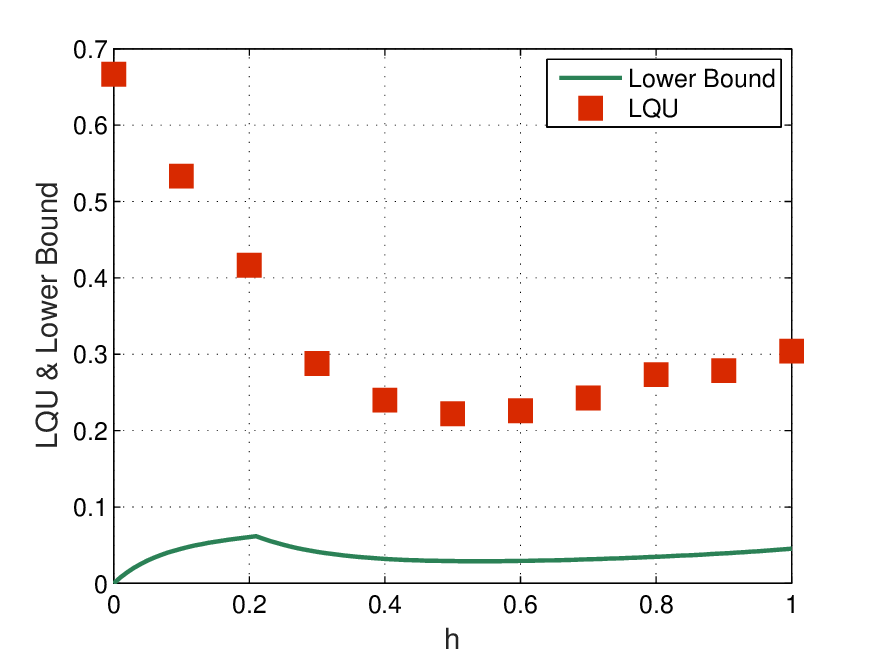}
\end{center}
\caption{The LQU of the $4 \times 2$ Horodecki state.}
\label{Horo42}
\end{figure}

We study the $4 \times 2$ Horodecki state \cite{horo42}
\begin{eqnarray}
{\rho_h}' = \frac{1}{{7h + 1}}\left( {\begin{array}{*{20}{c}}
h&0&0&0&0&h&0&0\\
0&h&0&0&0&0&h&0\\
0&0&h&0&0&0&0&h\\
0&0&0&h&0&0&0&0\\
0&0&0&0&{\frac{{(1 + h)}}{2}}&0&0&{\frac{{\sqrt {1 - {h^2}} }}{2}}\\
h&0&0&0&0&h&0&0\\
0&h&0&0&0&0&h&0\\
0&0&h&0&{\frac{{\sqrt {1 - {h^2}} }}{2}}&0&0&{\frac{{(1 + h)}}{2}}
\end{array}} \right).
\end{eqnarray}
It is also a PPT entangled state and does not satisfy ${\rm Tr}(\rho_h' \lambda_i \otimes \mathbb{I}_4)=0$. The procedure of obtaining the optimized LQU using the genetic algorithm is considerably lengthy. In the case of high-dimensional quantum systems, the operators tend to be degenerate. Thus the lower bound given Fig. \ref{Horo42} is much lower than the optimized LQU. However, when $h>0.2$, our bound still shows the similar trend as the optimized LQU.

\section{Discussion and Summary}

The importance of quantum states with higher dimensions (qudits) is gradually recognized in recent years. Compared with qubits, maximally entangled
qudits violate local realism more strongly and are less affected by noise
\cite{kaszlikowski2000,jchen2001,collins2002,cheng2009,son2006,he2011,ahrens2013}. In quantum communication, entangled qudits are
more secure against eavesdropping attacks \cite{bechmann2000,bourennane2001,cerf2002,durt2003,fpan2006},
and also offers advantages
including greater channel capacity for quantum communication \cite{fujiwara2003,liurong2011}
as well as more reliable quantum information processing \cite{ralph2007,fengli2011,smfei2013}.
Experimentally, the entangled qudits can be physically realized in linear photon systems \cite{moreva2006}, nitrogen-vacancy centres \cite{nv}, etc.
Therefore, it is urgent to establish a theory for measuring the quantum correlations in high-dimensional quantum systems.

Choosing an appropriate spectrum is crutial in the calculation of the LQU. Our lower bound is obtained by relaxation of the non-degeneracy fixed-spectrum requirement in the LQU definition. We have shown with this operation relaxation, the lower bound of the LQU is possible to be obtained.

For three-dimensional quantum systems, the freedom of the operator spectrum selection is relatively small. In this case, the lower bound is tight comparing to the optimized LQU. As the dimension of the quantum system grows, the freedom of the operator spectrum become larger. Although our bound turns to be much lower than the optimized value obtained with a specific non-degeneracy fixed-spectrum, it still reveals the trend of the LQU in the case where $h > 0.2$.

An interesting fact is that for the $4 \times 2$ Horodecki state, there exists a transition of the lower bound near $h = 0.22$. It has been shown the algebraic lower bound of the concurrence declines near $h = 0.22$, and decreases to zero when $h = 1$ \cite{physrep}, while the lower bound bounces back at $h = 0.55$, and is non-zero when $h = 1$. Our lower bound behaves similar to the lower bound of the concurrence.

Further research still needs to be done on both improving the lower bound and discovering more physical properties of high-dimensional quantum systems by using LQU.

\section*{Acknowledgements}

We thank Gerardo Adesso and Tommaso Tufarelli for helpful comments. We also thank Chenglong You, Debasis Sarkar, and Ajoy Sen for helpful discussions.
This work was supported by the National Natural Science Foundation
of China under Grant No. 11175094 and 91221205, the National
Basic Research Program of China under Grants No. 2015CB921002.

\end{document}